# Hybrid Semantic Tool Discovery for Enterprise MCP Gateway: Architecture and Implementation

**Olympia Saha**

**Amy Wang**

**Srinivasan Manoharan**

*PayPal, Inc.*

## Abstract

Large language model (LLM) agents invoke external tools to retrieve and reason over information beyond pretrained knowledge. The Model Context Protocol (MCP) standardizes how such tools are surfaced through a uniform interface, and a proxy MCP server aggregates many backend servers behind a single endpoint providing the enterprise a secure, governable chokepoint for authentication, policy enforcement, and observability. This architecture, however, creates two compounding challenges. First, a context-engineering bottleneck: when tools span many servers, their full schemas are loaded at the start of every conversation, saturating the model's context window before any user query and increasing inference cost while degrading tool-selection accuracy. Second, a user-facing tool discoverability barrier: with more than 200 MCP servers and 2,000+ indexed tools, users and agents cannot practicably identify which server or tool best addresses a given task through manual catalog inspection. Prompt caching reduces reprocessing cost but neither frees context capacity nor improves accuracy, since the model still attends to the entire catalog on every query. We present SCOUT (Selective Context Optimization for Universal Tooling), which applies context engineering principles, reframing tool exposure as a context-selection problem, injecting only the tools relevant to the current step. SCOUT surfaces two synthetic MCP meta-tools, tool_search and execute_tool, where tool_search performs agentic RAG via hybrid retrieval, fusing BM25 sparse matching with dense vector search through Reciprocal Rank Fusion to return the top-k relevant tools. Backed by coordinated ingestion, refresh, and retrieval pipelines supporting zero-downtime catalog updates, SCOUT resolves both the context saturation problem and the tool discovery challenge, enabling efficient, scalable tool access across enterprise environments. In production deployment at PayPal, SCOUT reduces MCP tool-token consumption from 140.2k tokens (70.1% of context) to 1.3k tokens (0.8%), a 99% reduction, directly cutting per-query inference cost at enterprise scale. Because SCOUT is surfaced as standard MCP tools, it is model-agnostic and requires no client-side modifications, making it immediately applicable to any MCP-compatible agent framework.

## 1 Introduction

Large language model (LLM) agents are increasingly deployed in enterprise settings where they must invoke external tools to retrieve real-time information, execute actions, and synthesize insights across heterogeneous data repositories. The Model Context Protocol (MCP) [1], introduced by Anthropic, provides a standardized JSON-RPC interface that connects LLMs to external data sources and services through a uniform tool-calling abstraction, enabling agents to invoke capabilities autonomously without bespoke integration code for each system. As MCP adoption accelerates across enterprise software ecosystems, tool catalogs will only grow larger, making scalable tool discovery an increasingly critical infrastructure challenge both for agents, which must select among thousands of candidate tool schemas per query, and for users, who cannot manually navigate a catalog of 200+ MCP servers to identify which capability best addresses their task.

In enterprise deployments, a proxy MCP server aggregates multiple backend MCP servers behind a single endpoint, presenting a unified tool namespace to agents while managing authentication, policy enforcement, and observability centrally. This architecture provides a secure and governable chokepoint for AI tool access at organizational scale. However, it introduces two compounding problems. The first is a context-engineering bottleneck: as the number of backend servers grows, the cumulative volume of tool schemas loaded at the start of every conversation rapidly saturates the model's context window before any user query has been processed, inflating inference costs and degrading tool-selection accuracy. Research has shown that LLM reasoning degrades when relevant information is buried among irrelevant content [9], a phenomenon that applies directly to tool selection when agents must choose among 2,000+ indexed tools. At PayPal, before SCOUT, every agent conversation began with 140.2k tokens already consumed by tool schemas, with 70% of the context window saturated before the first user message was processed. The second is a user-facing tool discoverability problem: with more than 200 MCP servers and 2,000+ tools in the catalog, users have no practical means of knowing which server hosts the capability they need or which specific tool to invoke. Manual catalog browsing is not viable at this scale. Natural language query is the only tractable interface for tool discovery in large enterprise catalogs.

Prompt caching partially alleviates the cost of reprocessing static tool schemas, but it does not reduce context window consumption or improve selection accuracy  the model still attends to the full tool catalog on every query regardless of relevance. What is needed is not cheaper loading of all tools, but selective loading of the right tools.

We present SCOUT (Selective Context Optimization for Universal Tooling), a context engineering system that reframes tool exposure as a context-selection problem. Rather than returning all 2,000+ tool schemas on every tools/list request, SCOUT surfaces two synthetic MCP meta-tools, tool_search and execute_tool, that decouple discovery from execution. An agent first calls tool_search with a natural-language query to retrieve only the most relevant tool schemas, then calls execute_tool to route the invocation to the appropriate backend server. Because SCOUT is implemented as standard MCP tools, it is model-agnostic and requires no client-side modifications, making it immediately applicable to any MCP-compatible agent framework. We have validated SCOUT across six MCP clients spanning three major providers: Claude Code, Claude Desktop, ChatGPT (Web and Desktop), GitHub Copilot (VS Code), Cursor, and OpenAI Codex CLI.

In production deployment at PayPal, SCOUT reduces MCP tool-token consumption from 140.2k tokens (70.1% of context) to 1.3k tokens (0.8%), a 99% reduction that directly cuts per-query inference cost at enterprise scale. The main contributions of this paper are: (1) a meta-tool protocol that resolves both user-facing tool discoverability and LLM context saturation, embedding semantic tool search natively within the MCP tool-calling flow so users can find the right tool across 200+ MCP servers through natural language rather than manual catalog browsing, without any client-side modifications; (2) a hybrid retrieval engine combining BM25 sparse matching and dense vector search with Reciprocal Rank Fusion, optimized for structured tool document representations; (3) user-scoped authorization filtering that enforces enterprise access control boundaries at the retrieval layer; (4) zero-downtime index management via an insert-before-delete upsert strategy with graceful degradation to full tool lists under infrastructure failure; (5) environment-aware dual-connector routing that enables users to direct tool execution across production and development environments through natural language intent alone, with no manual toggling, client-side configuration, or protocol extensions required; and (6) validated deployment across six MCP clients spanning three major providers, Claude Code, Claude Desktop, ChatGPT (Web and Desktop), GitHub Copilot (VS Code), Cursor, and OpenAI Codex CLI, without any client-side modifications.
The remainder of this paper is organized as follows. Section 2 surveys related work. Section 3 describes the AI Proxy platform. Section 4 formalizes the problem. Section 5 presents the system architecture. Section 6 details the implementation. Section 7 discusses broader implications and limitations. Section 8 concludes.

## 2 Related Work

### 2.1 Tool-Augmented Language Models

Tool use in LLMs was pioneered by Toolformer [2], which learned to self-supervise API call insertion. The ReAct [3] framework introduced the interleaved reasoning-and-action paradigm now standard in agentic systems. These approaches presuppose a small, fixed toolset provided at inference time. AutoGPT [4] and ReWOO [5] extended tool orchestration to multi-step planning but still load all tools into context. None addresses the scalability challenge that arises when the tool catalog grows to hundreds or thousands of tools across an enterprise ecosystem. ToolLLM [10] benchmarks tool use across 16,000+ real-world APIs, demonstrating that selection accuracy degrades significantly as catalog size grows. Gorilla [11] fine-tunes LLMs to select correct APIs from large catalogs but requires retraining for each new tool set. SCOUT addresses both limitations by operating at the retrieval layer, leaving the underlying LLM unchanged and adapting dynamically to evolving tool catalogs.

### 2.2 Retrieval-Augmented Generation

Retrieval Augmented Generation (RAG) [6] showed that dense passage retrieval can efficiently provide relevant knowledge to LLMs at inference time without retraining. Dense Passage Retrieval (DPR) [14] established the dual-encoder paradigm enabling retrieval based on meaning rather than surface-level term overlap. The BEIR benchmark [15] demonstrated that BM25 remains a competitive baseline across diverse retrieval tasks, often matching dense retrieval in zero-shot settings  a finding that directly motivates our hybrid design choice. Subsequent work explored hybrid retrieval combining BM25 sparse term matching with dense vector search to gain complementary recall [12]. Reciprocal Rank Fusion [7] provides a parameter free method for merging ranked lists from heterogeneous retrieval systems. While these techniques were developed for document retrieval, our work adapts them to a structurally distinct domain executable tool schemas  where parameter names, types, and descriptions serve as semantic anchors rather than free text. As Lumer and Subbiah [13] demonstrate in concurrent work, hybrid retrieval consistently outperforms either modality alone in agentic retrieval settings; SCOUT applies this insight specifically to the structured tool domain, where documents are parameter rich executable schemas rather than free text.

### 2.3 Vector Databases

Several vector databases support approximate nearest neighbor search at scale, including Pinecone, Weaviate, and Qdrant. We use Milvus [8], which provides native BM25 sparse vector support alongside dense HNSW indexing within a single query interface, enabling hybrid search without external preprocessing pipelines. Its built-in RRFRanker fuses sparse and dense ranked lists in-database, eliminating the need for an external merging step. As an open-source, self-hostable system, Milvus is also well-suited to enterprise environments with data residency requirements.

### 2.4 Retrieval in Agentic Systems

Retrieval in agentic systems differs fundamentally from standalone pipelines: rather than matching a fixed query against an index once, agents iteratively decide what to search, how many queries to issue, and whether results are sufficient or require refinement [13]. Lumer and Subbiah [13] provide the most directly related empirical study, evaluating lexical and semantic retrieval strategies across architecturally distinct agent harnesses including Claude Code, OpenAI Codex CLI, and Gemini CLI, the same clients validated in this work and finding that retrieval effectiveness varies significantly across harness classes even when the underlying corpus is held fixed. Their study focuses on document corpora (LongMemEval benchmark); SCOUT addresses a complementary problem retrieving from a catalog of executable tool schemas rather than free-text documents. Tool schemas present unique retrieval challenges: they are highly structured, parameter-rich, and domain-specific, requiring a representation that captures both the semantic intent of a tool and the precise vocabulary of its API surface. SCOUT's hybrid retrieval engine and structured tool document format are designed specifically for this domain. Their central finding that harness design influences retrieval effectiveness as much as the retrieval method itself directly motivates SCOUT's approach of embedding tool discovery as a first-class MCP meta-tool rather than an external preprocessing layer.

## 3 Background: The AI Proxy Platform

PayPal's AI Proxy is a FastAPI-based middleware service that provides a single governed interface to a growing ecosystem of internal MCP servers. It solves three organizational problems: tool aggregation (combining tools from disparate services into coherent proxy endpoints), access control (managing per-user and per-proxy server authorization), and structured logging for audit and governance.

### 3.1 Core Data Model

The platform organizes resources into three primary entities. An MCP Server represents a single upstream MCP-compliant service, described by a URL, transport type (SSE or streamable HTTP), and authentication configuration. A Proxy Server is a virtual aggregation endpoint that maps to one or more MCP Servers. Clients interact only with Proxy Servers, receiving a unified tool namespace spanning all constituent backends. A User Proxy MCP Server Mapping optionally enables an individual user to a subset of the MCP Servers within a given Proxy Server, enabling fine-grained access control. At PayPal, the platform spans 100+ internal MCP servers surfacing 2,000+ tools across domains including data analytics, engineering infrastructure, customer operations, and developer productivity.

### 3.2 Tool Caching

To avoid hammering upstream services, the platform employs a three-tier caching hierarchy: an in-memory cache (MemoryCache) for sub-millisecond lookups, a Redis cache for shared cross-process state, and a MySQL tool_cache table with a 24-hour TTL. This cache serves as the ground-truth tool registry from which the semantic index is populated. This layered architecture (live MCP connections for execution, multi-tier cached schemas for discovery) decouples retrieval latency from upstream server response time.

## 4 Problem Formulation

### 4.1 Formal Problem Setup

Let $T = \{t_1, t_2, ..., t_n\}$ denote a tool corpus of N tools managed by the AI Proxy, where each tool $t_i$ is described by a schema $\sigma(t_i)$ encoding its name, description, and parameter specification. In the PayPal deployment, N = 2,000+ tools span 100+ MCP servers. Each tool schema $\sigma(t_i)$ consumes $c(t_i)$ tokens when injected into the LLM context window. Given a user query q expressed in natural language, the tool discovery problem is to select a subset $S \subseteq T$, $|S| = k \ll N$, that maximizes relevance to q, subject to authorization constraints: for a user u with server permissions P(u), only tools from authorized servers may appear in S. The goal is to satisfy this retrieval objective while remaining within the context budget of the LLM and maintaining low end-to-end latency. A complicating factor is that tool names are not globally unique: two MCP servers may independently expose tools with the same logical name (e.g., search or query). The system therefore assigns each tool an enhanced name $\hat{e}(t_i)$ = {server($t_i$)}_{name($t_i$)}, where server($t_i$) is the sanitized MCP server name. When multiple server prefixes match, a longest-prefix-wins rule resolves ambiguity deterministically. The parameter k is user-configurable and clamped to [1, 20] with a default of 5, letting callers tune the recall–context-budget tradeoff at runtime without redeploying the system.

### 4.2 Why Full Tool Injection Fails

The naive approach of injecting all N tool schemas at the start of every conversation fails for three compounding reasons. First, context saturation: in the PayPal deployment, the aggregate token cost of all tool schemas is 140,200 tokens, consuming 70.1% of a 200,000-token context window before any user query is processed, leaving insufficient capacity for conversation history, system instructions, and agent reasoning. Second, accuracy degradation: Liu et al. [9] demonstrate that LLM reasoning degrades when relevant information is buried among irrelevant content, the 'lost in the middle' phenomenon. For tool selection, this means the model is more likely to invoke the wrong tool when the relevant schema is surrounded by 1,721+ irrelevant ones. Third, user-facing opacity: with more than 200 MCP servers and 2,000+ indexed tools, users cannot practicably identify the correct server or tool through manual catalog inspection. There is no directory or search interface in the base MCP protocol; users must either know the exact tool name in advance or browse the full tools/list response, neither of which scales beyond a handful of servers. Natural language query is the only tractable discovery interface at enterprise catalog scale, and full tool injection provides no support for it. Prompt caching partially addresses the first failure by amortizing repeated encoding costs but does not reduce context window consumption, mitigate accuracy degradation, or solve user-facing discoverability, as the model still attends to the full catalog on every forward pass.

### 4.3 Design Requirements

SCOUT must satisfy seven requirements simultaneously: (1) Recall-the relevant tool(s) for a given query must appear in the top-k results with high probability; (2) Context efficiency-tool-token consumption must be reduced from O(N) to O(k); (3) Latency-end-to-end tool search response must remain below 500 ms (p95) to preserve agent responsiveness; (4) Authorization-retrieved tools must respect per-user server access boundaries enforced at the retrieval layer, not as a post-processing filter; (5) Catalog freshness-the tool index must support zero-downtime updates as MCP server tool registries evolve; (6) Client agnosticism-the system must operate transparently within the standard MCP protocol, requiring no client-side modifications, so that any MCP-compliant client (Claude Code, Claude Desktop, ChatGPT, GitHub Copilot, Cursor, OpenAI Codex CLI) can adopt it without integration work; and (7) Graceful degradation-when retrieval infrastructure is partially or fully unavailable, the system must degrade gracefully rather than fail, maintaining MCP protocol compatibility throughout. No prior approach satisfies all seven simultaneously: full injection violates (2); dense-only retrieval struggles with (1) for exact API name queries; Gorilla-style fine-tuning [11] violates both (5) and (6) as it requires retraining when tools change and assumes client awareness of the retrieval step; and none address (4) at the retrieval layer; and all assume reliable retrieval infrastructure, violating (7). SCOUT addresses (7) through a two-stage fallback: when the OpenAI embedding call fails, it degrades to BM25-only sparse search served directly from Milvus; when the retrieval index is entirely absent, it falls back to all the tools injection available through the proxy preserving MCP protocol compatibility at the cost of temporary context efficiency loss.

## 5 System Architecture

The SCOUT architecture comprises five interacting subsystems: the Tool Ingestion Pipeline, which converts MCP server tool schemas into searchable vector documents; the Hybrid Index (Milvus), which stores and retrieves tools using combined BM25 sparse and dense vector search; the tool search meta-tool, which receives natural-language queries and returns ranked tool schemas; the execute tool meta-tool, which routes tool calls to the appropriate upstream MCP server; and the Index Lifecycle Manager, which keeps the index synchronized with server health and catalog changes. Figure 1 illustrates the end-to-end data flow across all three operational paths.

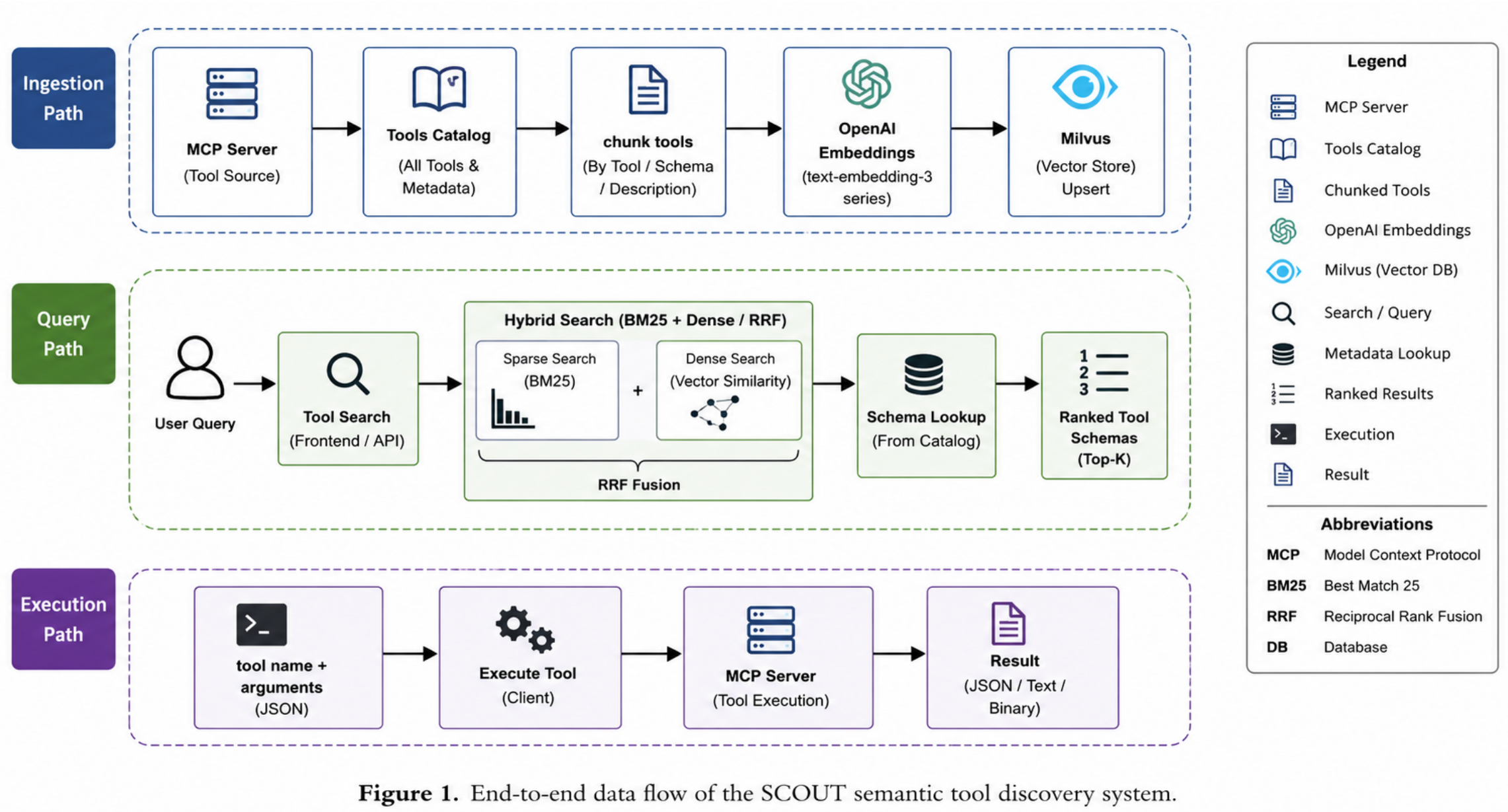

**Figure 1.** End-to-end data flow of the SCOUT semantic tool discovery system.

### 5.1 Two-Phase Query Protocol

When an MCP client issues a tools/list request to the AI Proxy, it receives a stub list containing only two tools: tool_search and execute_tool. The full catalog-2,000+ tool schemas in the PayPal deployment is never transmitted to the client. The LLM's two-phase behavior is orchestrated entirely through the tool description text, requiring no special fine-tuning, protocol extensions, or client-side modifications. The tool_search description is labeled "REQUIRED FIRST STEP" and enumerates the PayPal domain categories it covers-People & Org, Internal Business Applications, Analytics & BI, Infrastructure & Observability, and Dev & Knowledge-so the client knows when to invoke it. It also embeds the mandatory workflow as instructions: call tool_search with a natural-language query, inspect the returned schemas, then call execute_tool. A hard guardrail is included: "Never guess a tool_name. Always get it from tool_search first." The execute_tool description is labeled "REQUIRED SECOND STEP" with the same guardrail: the tool_name must be the exact string returned by tool_search. In phase one, the proxy performs hybrid semantic retrieval and returns the schemas of the top-k most relevant tools (default k=5, configurable to [1, 20]). In phase two, the proxy routes the call to the appropriate upstream MCP server. This two-phase protocol reduces tool-definition token consumption from 140,200 tokens to approximately 1,300 tokens - a 99% reduction while preserving full tool execution capability across all 2,000+ tools. A key efficiency property emerges from the description design: the execute_tool description mirrors the same domain categories as tool_search - People & Org, Analytics & BI, Infrastructure & Observability, and Dev & Knowledge giving both meta-tools overlapping semantic fingerprints. As a result, any query that retrieves tool_search will also retrieve execute_tool in the same top-k result set. The MCP client therefore receives both tool schemas in a single search, with no second lookup needed to discover the execution mechanism.

This is reinforced explicitly in the tool_search description: "After calling tool_search you can immediately call execute_tool you already have its full schema above. Do NOT call tool_search again to find execute_tool." Because the entire protocol is expressed through standard MCP tool descriptions, it works identically across all validated clients: Claude Code, Claude Desktop, ChatGPT, GitHub Copilot, Cursor, and OpenAI Codex CLI. This description-driven orchestration principle also extends naturally to multi-connector deployments, where separate production and development connectors each surface their own tool_search and execute_tool pair; environment selection is handled through natural language intent alone, as described in Section 5.7.

### 5.2 Conditional Meta-Tool Injection

SCOUT can be enabled or disabled independently at two levels, giving operators fine-grained control over rollout. At the client level, the tool_search_clients configuration lists the MCP client types that receive meta-tools instead of the full tool catalog (e.g., ["claude-code"]). Adding a client name here enables SCOUT globally for that client type across all proxies. At the proxy level, the tool_search_proxy_ids configuration lists specific proxy UUIDs that receive meta-tools when the connecting client has no registered name (e.g., openai-mcp clients or direct API integrations). Adding or removing a proxy ID from this list is the per-proxy on/off switch, enabling operators to roll out SCOUT to selected proxy endpoints without affecting others. Meta-tool injection occurs only when three conditions are all satisfied: (1) the proxy being accessed is listed in tool_search_proxy_ids, or the connecting client's name appears in tool_search_clients together these two toggles determine whether the proxy is SCOUT-enabled for this client; (2) the RAG index reports itself as initialized and non-empty. This conditional injection ensures that standard proxy deployments are completely unaffected, that clients not yet enabled for tool_search receive the conventional tools/list response with full tool schemas behind the proxy, and that the meta-tools are never exposed when the backing index is unavailable.

### 5.3 Index Lifecycle Management

The Milvus index is health-aware and status-driven: only tools from ACTIVE servers exist in the index at any time. MCP servers transition through five states: PENDING, ACTIVE, DEGRADED, INACTIVE, and REJECTED - each triggering a corresponding Milvus action. When an administrator approves a PENDING server (PENDING → ACTIVE) into the gateway, its tools are immediately ingested into Milvus and become searchable. When the health check service detects accumulated connection failures (ACTIVE → DEGRADED), the server's chunks are deleted from Milvus so degraded tools never surface in search results. Recovery is handled by scheduled job. It tests each DEGRADED server past its cooldown period via a lightweight metadata endpoint probe, and on success transitions the server back to ACTIVE and triggers re-ingestion. Manual deactivation or rejection (→ INACTIVE or REJECTED) likewise removes the server's chunks from Milvus. A separate scheduled job polls the catalog for any MCP server modified in a rolling window. For each matched server, it clears the DB tool cache, invalidates Redis and in-memory caches, re-fetches tool schemas from the upstream server, and re-ingests into Milvus ensuring the index reflects the latest tool definitions across the enterprise catalog without manual intervention.

### 5.4 Scope-Based Search Filter

Tool search supports two search filters. In all_servers' mode (the default), tool search searches all ACTIVE MCP servers in the index. In my_servers' mode, SCOUT identifies which MCP servers the requesting user has explicitly added to their proxy and restricts retrieval to that subset. In both modes, the server names are translated into a Milvus metadata filter applied at query time so that search is enforced inside the vector database rather than as a post-retrieval filter.

### 5.5 Ingestion Pipeline

The ingestion pipeline converts raw MCP server tool schemas into searchable vector documents in Milvus. It is triggered by three events: (1) a server transitions to ACTIVE status (on initial approval or recovery from DEGRADED); (2) an administrator manually triggers a tool schema retrieval; or (3) if scheduled job detects that a server's definition has changed in the enterprise catalog. In all cases the pipeline follows the same four stages. First, tool schemas are loaded from the three-tier cache (in-memory → Redis → MySQL tool_cache table), which decouples ingestion latency from upstream MCP server availability. Second, tools are chunked, converting each schema into a structured natural-language document that interleaves the tool name, server category, description, and parameter specifications in a format optimized for both keyword and semantic retrieval. Third, the documents are embedded in batches using OpenAI text-embedding-3-large, producing 3,072-dimensional dense vectors. Fourth, chunks and embeddings are upserted into Milvus using an insert-before-delete strategy: new vectors are written first, then stale vectors for the same server are deleted, ensuring zero search downtime during updates. Each chunk carries an entity_source field in the format {server_name}.tool, which serves as both the filter key for tool search and the deletion scope for lifecycle events.

### 5.6 Retrieval Pipeline

The retrieval pipeline executes on every tool_search call and comprises four stages. First, the gateway resolves the requesting user's session to determine the client’s name and user identity, then applies the three-way injection gate (Section 5.2) to confirm the request should be handled by SCOUT. Second, the filter scope is resolved based on in my_servers’ or all_servers’ mode. The mode is translated into a Milvus metadata filter that enforces filtering inside the vector database. Third, the HybridIndex executes a two-branch search against Milvus: the dense branch embeds the query with text-embedding-3-large and retrieves the top-k×3 nearest neighbors by cosine similarity; the sparse branch runs BM25 over the same query against the Milvus-native sparse index. Each branch deliberately fetches three times the requested results - a candidate expansion strategy that ensures documents ranking highly in either branch are not missed by the fusion step (see Section 6.4 for details). The two ranked lists are fused with Reciprocal Rank Fusion (k=60), and the final top-k results are returned. If the OpenAI embedding call fails, the pipeline degrades to BM25-only retrieval. Fourth, schema hydration maps each retrieved chunk back to the full tool schema stored in the cache, assembles the ranked response, and returns it as a standard MCP JSON-RPC result.

### 5.7 Dual Connector Support and Environment-Aware Routing

When both a production and a development (or non-production) connector are registered simultaneously in an MCP client such as Claude Desktop, SCOUT surfaces two independent pairs of meta-tools: one tool_search and execute_tool bound to the production environment and a second pair bound to the development environment. This multi-connector topology enables environment-aware tool execution driven entirely by natural language intent, with no manual toggling or client-side configuration required from the user. Each connector's tool_search description explicitly identifies its target execution environment. The production connector labels itself accordingly in its description text (for example, 'production environment' or 'live data'), while the development connector identifies itself as development, staging, or non-production. When a user expresses production intent, for example 'I want to query production data' or 'run this against the live environment,' the LLM selects the production connector's tool_search and execute_tool pair. A request framed around sandbox, staging, or development data routes through the development connector pair. The same semantic matching that selects the right tool within a single connector also selects the right connector across environments. Crucially, if a capability such as a BigQuery MCP server exists in both environments, the user does not need to know which connector hosts which server. Stating 'query production

BigQuery' is sufficient for the LLM to resolve both the environment and the specific tool, issuing a single tool_search against the correct connector and then executing via that connector's execute_tool. This design extends SCOUT's description-driven orchestration principle: the same mechanism that drives the two-phase query protocol, using tool description text as a behavioral specification, handles multi-environment dispatch without additional protocol machinery, client modifications, or explicit routing configuration.

## 6 Implementation Details

### 6.1 Tool Document Construction

Each tool is represented as a structured natural-language document for embedding. The ingestion module constructs this document by combining the tool name, its MCP server category, the MCP server description, the tool description, and a human-readable rendering of each parameter including type, enum values where present, and required/optional status. Including the server description provides broader domain context that anchors retrieval when a user query describes the server's area of expertise rather than a specific tool name, improving recall for cross-server disambiguation. This structured representation provides multiple semantic anchors, the tool name, server context, description, and parameter semantics, maximizing the probability that a user's natural-language query will match the embedding of the correct tool.

```
Tool: {tool_name} Category: {mcp_server_name} Server Description: {mcp_server_description}
Description: {tool_description} Parameters:  - {param_name} [{type}](required/optional):
{param_description}
```

Each chunk is assigned a deterministic ID in the form {server_name}_{tool_name}, enabling idempotent upserts. The entity_source field stores the server’s name as {server_name}.tool, which is used as the Milvus filter expression during retrieval.

### 6.2 Milvus Collection Schema

The Milvus collection stores seven fields per chunk: chunk_id (primary key, VARCHAR), text (VARCHAR with BM25 analyzer enabled), embedding (FLOAT_VECTOR of dimension 3,072), sparse (SPARSE_FLOAT_VECTOR generated automatically by the BM25 built-in function), entity_name (bare tool name), entity_source (server name), and chunk_json (full JSON serialization of the chunk for result hydration). Two indexes are maintained: an HNSW index on the dense embedding field with COSINE metric, and a SPARSE_INVERTED_INDEX on the BM25 sparse field.

### 6.3 Embedding Strategy

Dense embeddings are generated using OpenAI's text-embedding-3-large model, which produces 3,072-dimensional vectors. This model was chosen for its superior semantic capture quality. Embedding calls are batched in groups of 16 to balance throughput and API rate limits, and a tenacity-based retry decorator provides automatic back-off on RateLimitError, APIConnectionError, and APITimeoutError, ensuring resilience against transient API failures.

### 6.4 Hybrid Retrieval with RRF Fusion

At query time, the hybrid search performs a two-branch retrieval. The dense branch embeds the query using the same text-embedding-3-large model and executes an AnnSearchRequest against the HNSW index with COSINE metric, retrieving top_k × 3 candidates to provide sufficient overlap for fusion. The sparse branch executes a BM25 search against the SPARSE_INVERTED_INDEX using the raw query text, also retrieving top_k × 3 candidates. The two result sets are fused by Milvus's native RRFRanker with k=60, producing a single ranked list of top_k results. RRF is parameter-light: scores are computed as:

$$RRF_score(d) = \Sigma_r \; 1 / (k + rank_r(d))$$

where the sum is over the BM25 and dense ranked lists and rank_r(d) is the rank of document d in list r. The top_k × 3 over-fetching per branch is a deliberate candidate expansion strategy. With top_k = 5, each branch independently retrieves 15 candidates before fusion. This is necessary because the two branches produce different orderings: a document that ranks 10th in the dense branch but 2nd in BM25 may be the best overall result after fusion but if each branch fetched only top_k = 5 results, that document would never reach the RRF ranker. Over-fetching gives the ranker a wider 30-candidate pool, so documents that score highly in either branch (but not both) are still considered in the final fusion. The tradeoff is a modest increase in Milvus retrieval work per query, offset by meaningfully improved result quality for queries where the two modalities disagree on ranking. The hybrid approach is motivated by the complementary strengths of the two modalities: BM25 reliably matches exact tool names and parameter keywords (e.g., "create_pull_request"), while dense retrieval handles paraphrases and intent-level queries (e.g., "open a new GitHub PR"). If the OpenAI embedding call fails due to an API outage, the system automatically falls back to BM25-only search, preserving partial functionality.

### 6.5 Meta-Tool Interface Specification

The tool_search meta-tool accepts three input parameters. The query parameter (string, required) carries the natural-language task description; its parameter description instructs the LLM to phrase it as a task rather than a tool name, since retrieval is semantic. The top_k parameter (integer, optional, default 5, clamped to [1, 20]) controls how many ranked tool schemas are returned, letting the caller trade recall breadth against context consumption. The scope parameter (string, optional, enum: my_servers | all_servers) restricts retrieval to the user's personally mapped servers or all active servers on the proxy; this parameter can be conditionally omitted from the schema for fixed-scope proxies. The tool_search response is returned as an MCP text-content block containing a JSON object with two fields: query (the echoed search string, enabling the LLM to verify what was retrieved) and matches (an ordered array of tool descriptors). Each match carries four fields: name (the enhanced tool name in {server}_{tool} format), server_name (the bare MCP server name), description, and inputSchema (the full JSON Schema for the tool's parameters). Wrapping the response as text rather than a structured MCP result is intentional: it lets the LLM read and reason over the ranked list as free text before committing to a tool call. In dual-connector deployments, each connector's tool_search description additionally carries an environment identifier (for example, 'production environment' or 'development environment') embedded in the tool description text. This label is the sole signal the LLM uses to route between environments, requiring no schema changes or additional parameters.

The execute_tool meta-tool accepts three required parameters. The tool_name parameter (string, required) must be the exact enhanced name returned by tool_search - its description includes the explicit guardrail "Do not guess." The server_name parameter (string, required) must be the exact server_name field from the tool_search match; the proxy uses this to route directly to the correct upstream MCP server. The arguments parameter (object, required) must conform to the inputSchema returned by tool_search for that tool. Both tool_name and server_name carry identical guardrail descriptions in their parameter definitions, reinforcing the constraint that both values must come verbatim from tool_search results and must never be hallucinated. The execute_tool response is the raw result from the upstream MCP server, passed through unmodified.

### 6.6 Observability

SCOUT instruments every stage of the two-phase query protocol with structured log events, enabling a Splunk-based observability dashboard that spans session entry through tool execution. Four monitoring tiers are derived from these events.

Session Entry Volume

The tools_list_tool_search event is emitted each time an MCP client calls tools/list and SCOUT is active for that proxy. Tracking this event over time reveals adoption trends and session volume per proxy server, surfacing which clients and proxies are driving the most SCOUT activity. Per-proxy breakdowns identify high-traffic endpoints and support capacity planning.

RAG Search Analytics

The tool_search_completed event carries structured fields including results_count, matched_servers, latency_ms, top_k, and scope. These fields power four classes of analysis. Result quality metrics track the distribution of results_count values and the fraction of searches returning zero results, broken down by proxy and query scope. MCP backend surfacing frequency identifies which MCP servers appear most often in search results via the matched_servers' field, revealing which upstream capabilities are most discoverable. Scope distribution tracks the ratio of my_servers to all_servers' invocations, characterizing user intent patterns. Latency distribution is monitored at p50, p95, and p99 with time-series trending to detect performance regressions as the index scales.

Fallback Rate Monitoring

The tool_search_fallback event is emitted at WARNING level when SCOUT falls back to full tool injection due to RAG unavailability either Milvus unreachable or the OpenAI embedding call failing. Fallback rate is computed as the ratio of tool_search_fallback to tool_search_completed events over a rolling window, providing a direct measure of RAG system reliability. Latency comparisons between the RAG path and the fallback path quantify the performance cost of degraded operation, motivating rapid remediation of embedding service disruptions.

Tool Execution Analytics

The execute_tool_completed and execute_tool_failed events enable downstream execution analysis. Most-invoked tool rankings identify high-demand capabilities and inform MCP server prioritization in the index. Per-server call volume tracks backend load distribution across MCP servers. Failure heatmaps tabulating error counts by tool_name and server_name pair, and by error class and error category pinpoint which tools or error types require attention, distinguishing transient network failures from systematic tool-level defects.

## 7 Evaluation

### 7.1 Retrieval Quality

We evaluate SCOUT's retrieval quality using a 49-query benchmark spanning nine domain categories (data analytics, observability/SRE, productivity, payments/merchant, developer tooling, risk/fraud, infrastructure, BI reporting, and operations). Queries were executed against the live production catalog of 2,000 indexed tools. Four queries expose catalog gaps (N/A): web page scraping (Q4), screenshots (Q10, Q50), and Confluence wiki access (Q37). The remaining 45 evaluable queries reflect intentionally diverse and challenging retrieval scenarios, including cross-domain disambiguation and queries targeting less-populated catalog regions.

Table 1 shows per-query results for the original 15-query subset (13 evaluable). Across the full 49-query benchmark (45 evaluable), SCOUT achieves Hit@1 of 84.8% (38/45), Hit@5 of 95.6% (43/45), and MRR of 0.821. The 35 newly added queries deliberately probe harder retrieval scenarios: Hit@1 on the expansion set is 75.6%, with correct tools still surfaced within the top 5 in 93.9% of evaluable cases. Four catalog gaps expose coverage boundaries: Confluence wiki documentation, screenshot capture, and general-purpose web scraping have no MCP server in the current hub. All other domains achieve at least one relevant tool in the top-5 results.

Query 5 ('send a Slack message to a channel') illustrates an important distinction from catalog gaps like Q4. PayPal does not have a Slack MCP server; Microsoft Teams is the enterprise messaging platform in deployment. The correct answer for this query therefore is teams_send_message, which exists in the catalog. A pure BM25 search returns zero matches because the term 'Slack' appears in no tool description. The dense branch correctly bridges this vocabulary gap, surfacing teams_send_message by mapping the channel messaging semantic intent to the available enterprise tool. This is a retrieval success (Y*), not a catalog gap: the right capability exists but under a different product name, and hybrid retrieval finds it where lexical search cannot.

**Table 1: Original 15-Query Spot-Check (13 Evaluable, Q4 and Q10 = N/A)**

| # | Query | Top-1 Retrieved Tool | H@1 | H@5 |
|---|---|---|---|---|
| 1 | Create Jira ticket for a bug | create_jira_ticket_for_failure | Y | Y |
| 2 | Get employee manager org chart | get_organizational_chart | Y | Y |
| 3 | Run BigQuery SQL analytics query | query_bigquery | Y | Y |
| 4 | Scrape webpage and extract text | (no tool in catalog) | N/A | N/A |
| 5 | Send a Slack message to a channel | teams_send_message | Y* | Y |
| 6 | Look up a Confluence/wiki article | get_article (domain KB) | ~ | ~ |
| 7 | Check Kubernetes pod health and logs | get_app_k8s_status | Y | Y |
| 8 | Get payment transaction metrics | pf_sndr_summary | Y | Y |
| 9 | Schedule a calendar meeting | respond_to_meeting (intent miss) | N | Y |
| 10 | Take a screenshot of a web page | (no server in catalog) | N/A | N/A |
| 11 | Find active production incidents | fetch_ongoing_incidents | Y | Y |
| 12 | Get a Looker dashboard or BI report | get_dashboard_details | Y | Y |
| 13 | Check batch job status and schedule | list_standin_batch_jobs | Y | Y |
| 14 | Analyze error logs and suggest fix | analyze_log_for_errors | Y | Y |

| 15 | Generate code from PR or template | approve_tests (workflow step) | N | Y |
|---|---|---|---|---|
|  | **Overall** |  | **84.6% (n=13)** | **100% (n=13)** |

Y = correct tool at rank 1; N = wrong tool at rank 1 but correct tool in top-5; ~ = partial match (no dedicated tool in catalog, closest analog returned); N/A = no tool of this type exists in catalog (Q4: no web scraper); Y* = correct tool found via cross-product semantic mapping (Q5: Slack query correctly resolved to Teams, the organizational messaging platform). Metrics computed over 14 queries excluding Q4. MRR = 0.929.

*Table 1b: SCOUT Hybrid Retrieval on Expanded 49-Query Benchmark (45 Evaluable, 4 N/A Catalog Gaps)*

| **Metric** | **Original 15-Q (n=13)** | **Expanded 35-Q (n=33)** | **Combined 49-Q (n=45)** |
|---|---|---|---|
| Hit@1 (Y + Y* + ~) | 84.6% | 75.6% | 84.8% |
| Hit@5 | 100% | 93.9% | 95.6% |
| MRR | 0.915 | 0.784 | 0.821 |
| N/A catalog gaps | 2 (Q4, Q10) | 2 (Q37, Q50) | 4 total |

Table 1b summarizes retrieval metrics across the full 49-query benchmark. Compared with the original 15-query spot-check (Hit@1 84.6%, Hit@5 100%), the expanded benchmark shows modestly lower Hit@1 (84.8% on all 45 evaluable) because the 35 new queries deliberately target harder and less-common retrieval scenarios. Hit@5 remains high at 95.6%, confirming that the hybrid retrieval consistently finds relevant tools within the returned candidate set even when ranking is imperfect.

Domain-level performance reveals that analytics and observability queries achieve near-perfect Hit@1 (DFS metrics server surfaces at rank 1 for 11 of 11 analytics queries). SRE and infrastructure queries achieve Hit@1 of 78% (7/9 evaluable), with failures concentrated on runbook retrieval (no dedicated runbook MCP server) and OOMKill-specific diagnosis (correct node-analyzer tool surfaces at rank 2). Productivity queries achieve Hit@1 of 75% (6/8 evaluable), with SharePoint document search and general Outlook email routing as failure cases. BI and reporting queries show the lowest Hit@1 (50%, 2/4 evaluable) because Looker dashboard tools surface at rank 4 when queries are phrased with payment-metric context rather than Looker-specific vocabulary.

Server-side processing latency, logged via the tool_search_completed structured event and measured from the start of hybrid retrieval through response serialization, averaged 572 ms (median: 440 ms, P95: 936 ms) across 1,921 production requests over a 24-hour window against the PayPal MCP Hub. This figure encompasses Milvus hybrid search (BM25 + dense vector + RRF), tool schema resolution from the database cache, and response formatting. The minimum observed latency of 232 ms represents the floor for a warm-cache retrieval against a 2,000-tool corpus.

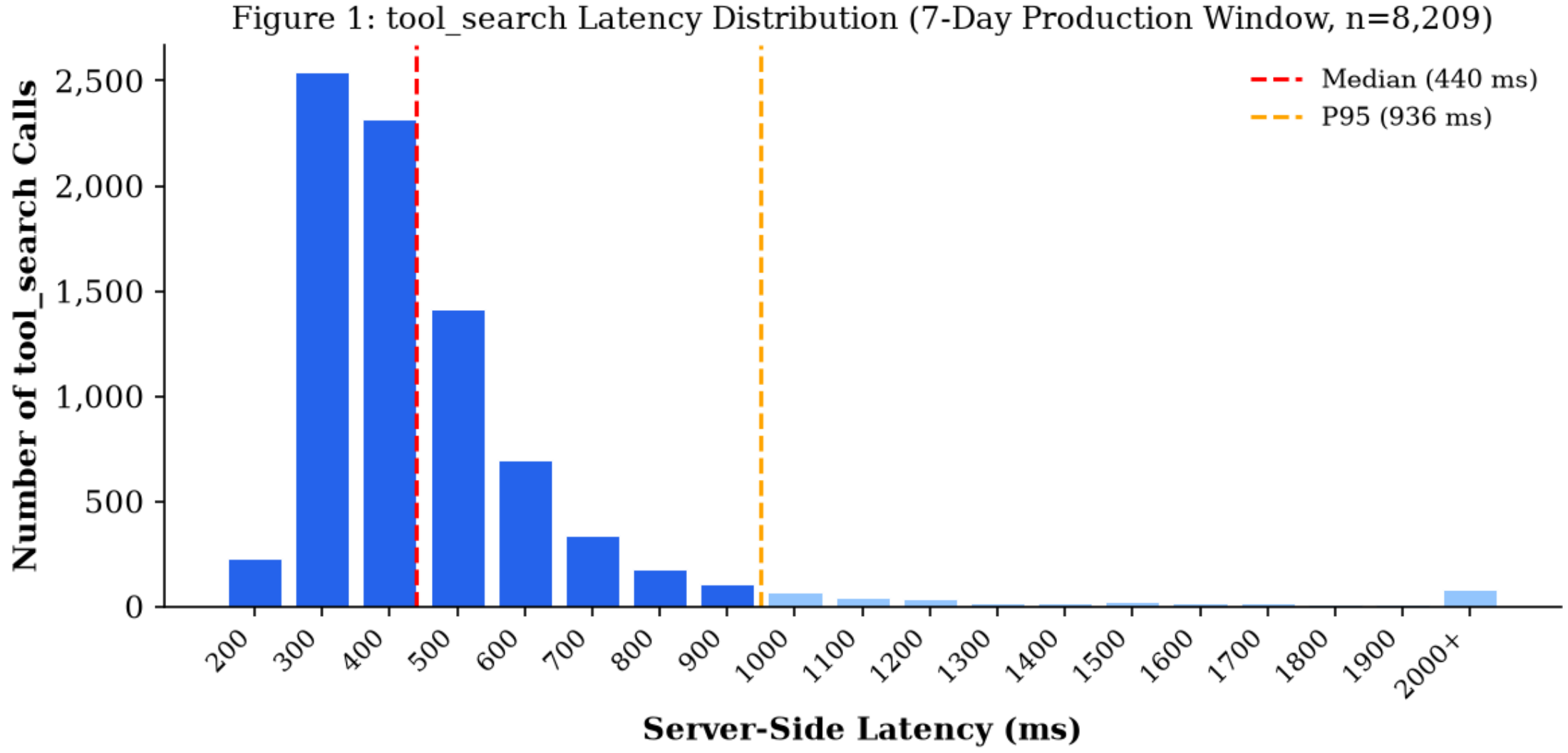

*Figure 1: Server-side tool_search latency distribution over a 7-day production window (n = 8,209). The distribution is right-skewed, with a mode at 300-400 ms and a long tail driven by cold-cache retrievals and peak-load contention.*

## 7.2 Context Efficiency

Table 2 summarizes context consumption before and after SCOUT deployment. Prior to SCOUT, every agent session began with 140,200 tokens consumed by tool schemas, representing 70.1% of a 200,000-token context window before the first user message. SCOUT reduces per-session context to 1,300 tokens (0.8%), a 99% reduction at production scale. In the spot-check evaluation, each query retrieved an average of approximately 2,500 tokens of tool schema (top-5 results), confirming that context overhead remains low even when the retrieval set is expanded beyond the production default.

**Table 2: Context Consumption Before and After SCOUT**

| Scenario | Tokens Injected | % of 200k Context |
|---|---|---|
| Full catalog, pre-SCOUT | 140,200 | 70.1% |
| SCOUT, production deployment | 1,300 | 0.8% |
| SCOUT, spot-check avg. (top-5) | **~2,500** | ~1.25% |

### 7.3 Failure Mode Analysis

Two failure modes were identified across the 15-query spot-check, each with distinct implications for system design.

Intra-domain intent disambiguation. Query 9 ('schedule a calendar meeting and send invites') is notable because all five retrieved tools were correct calendar tools from the same server (create_event, respond_to_meeting, get_free_busy, delete_event, update_event), demonstrating accurate domain identification. The failure is purely in intra-domain ranking: respond_to_meeting ranked first because its description contains 'scheduling workflows' and 'send_response' language that lexically overlaps with 'schedule' and 'send invites' in the query. The correct tool create_event was recovered at rank 2 (Hit@5 = Y). An intent-classification pre-filter that resolves action type ('create' vs 'respond' vs 'cancel') before retrieval would address this class of failure.

Workflow step confusion. Query 15 ('generate code from a pull request or template') surfaced approve_tests at rank 1, an intermediate workflow step, rather than the entry-point generate_pr tool. Multi-step workflows expose each step as an independent MCP tool; without workflow-level provenance, retrieval cannot distinguish entry points from sub-steps. A workflow-aware chunking strategy, grouping related steps under a single entry-point document, is a natural direction for future work.

### 7.4 Production Reliability

Table 3 reports operational metrics drawn from the AI Gateway structured log index in Splunk over a 24-hour production window. These figures reflect real user activity on the PayPal MCP Hub and provide evidence of SCOUT's reliability and adoption at enterprise scale.

Table 3: SCOUT Production Operational Metrics (24-Hour Window, PayPal MCP Hub)

| Metric | Value (24-Hour Window) |
|---|---|
| MCP sessions initiated | 31,701 |
| Total MCP tool calls | 55,314 |
| tool_search invocations | 1,926 |
| Avg. tools returned per search | 5.5 (median: 5) |
| execute_tool completions | 8,694 |
| tool_search fallback rate | 0.0% (Milvus always available) |
| tool_search invocation rate | 43.1% (1,926 / 4,464 exposures) |
| Search-to-execution ratio | 1 : 4.5 |

Several observations merit attention. First, the zero fallback rate confirms that the Milvus hybrid index is available at production reliability levels, with no degradation to full-catalog injection during the observation window. Second, the 1:4.5 search-to-execution ratio reflects the intended SCOUT workflow: a single tool_search call surfaces the relevant tool set, after which the LLM issues multiple execute_tool calls within the same session. Third, the 43.1% tool_search invocation rate (searches per tools/list exposure) indicates that LLMs selectively invoke tool_search when needed rather than on every turn, consistent with the meta-tool protocol design.

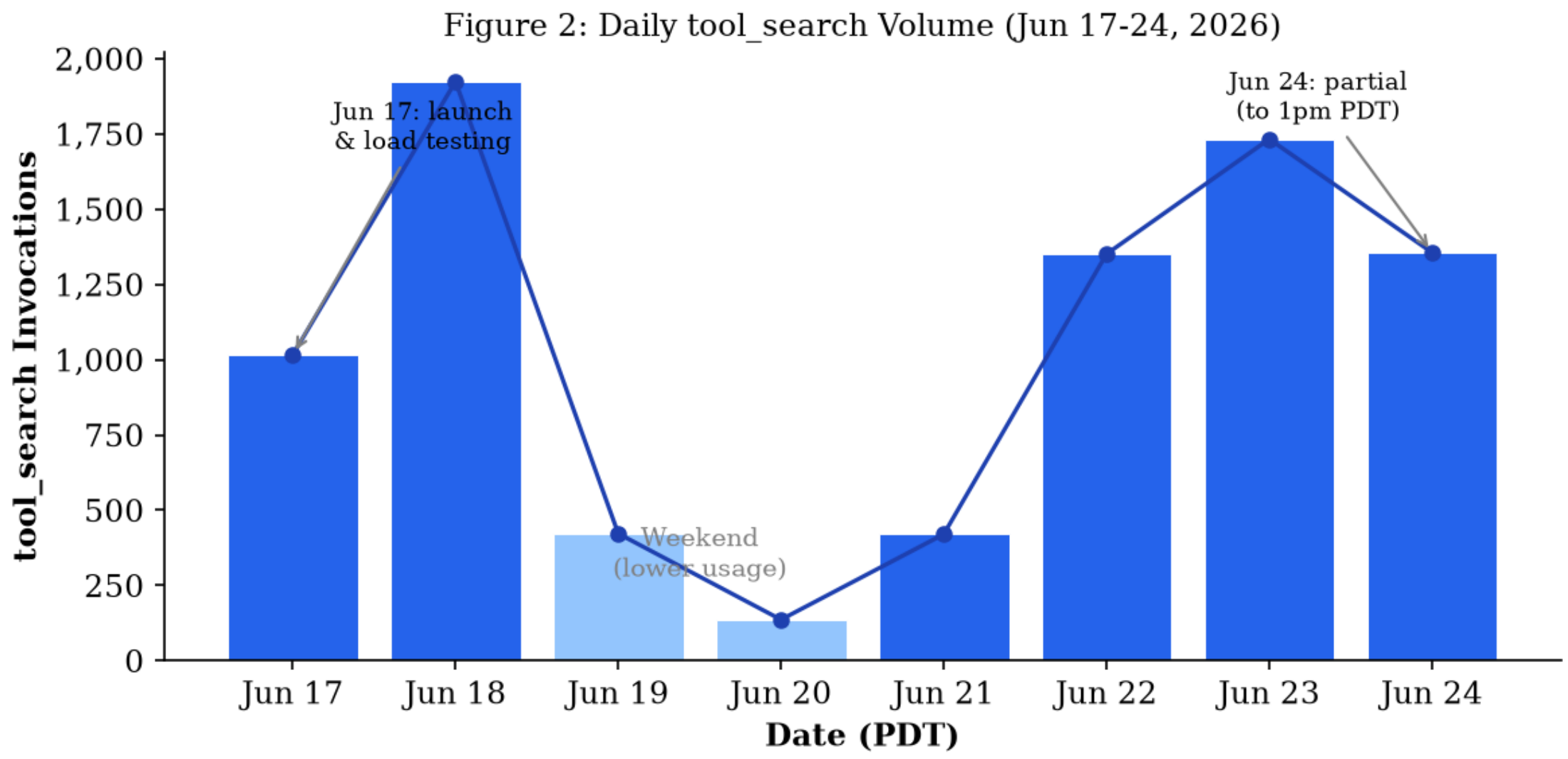

*Figure 2: Daily tool_search invocation volume (Jun 17-24, 2026). Jun 17 reflects launch-day load testing. The weekend dip (Jun 19-20) and subsequent weekday recovery confirm that tool_search usage follows enterprise business-hours patterns.*

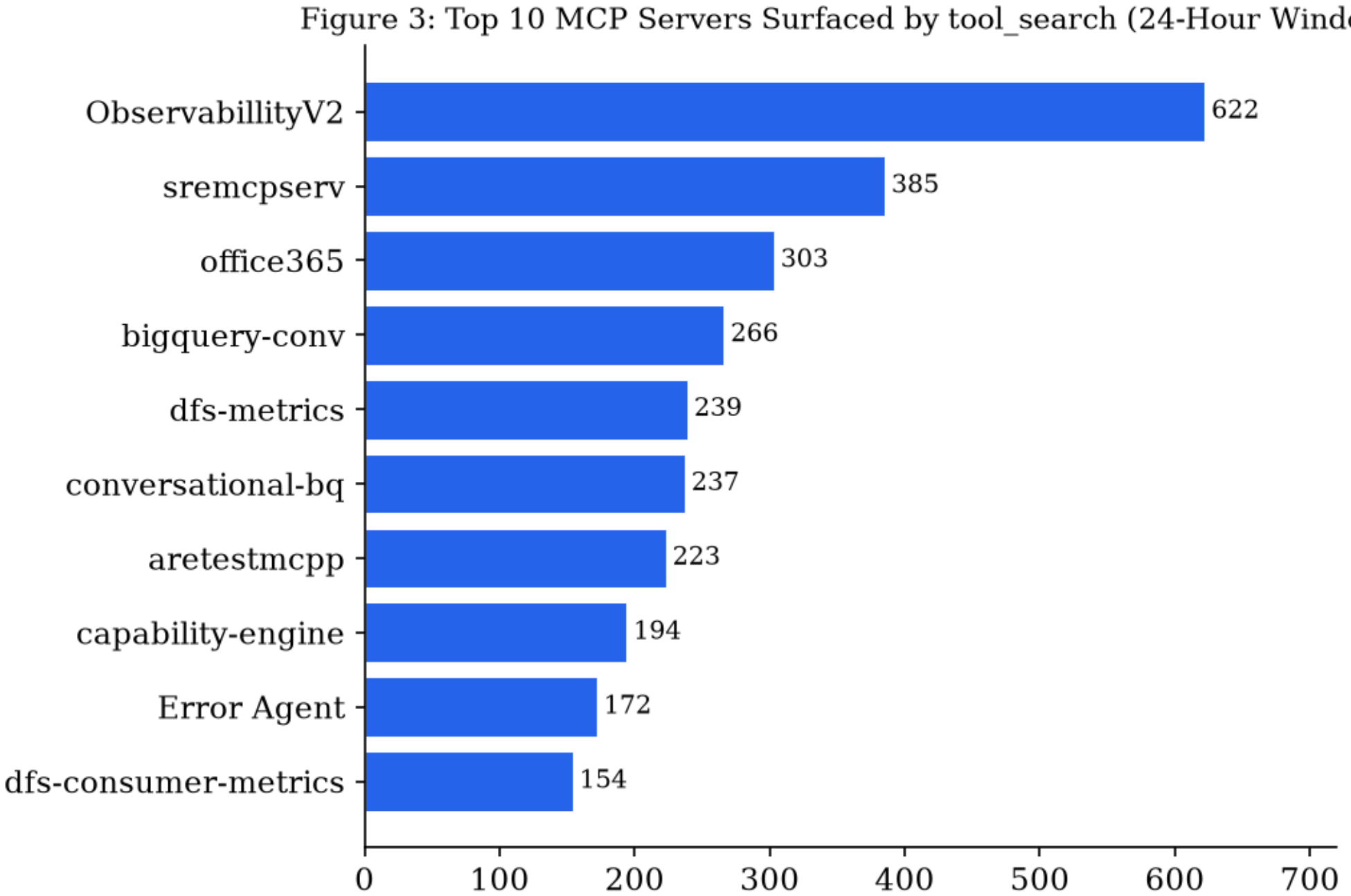

*Figure 3: Top 10 MCP servers surfaced by tool_search in the 24-hour observation window. The breadth of server categories, spanning observability, SRE, productivity, analytics, and domain-specific agents, demonstrates that SCOUT routes queries across the full enterprise tool catalog.*

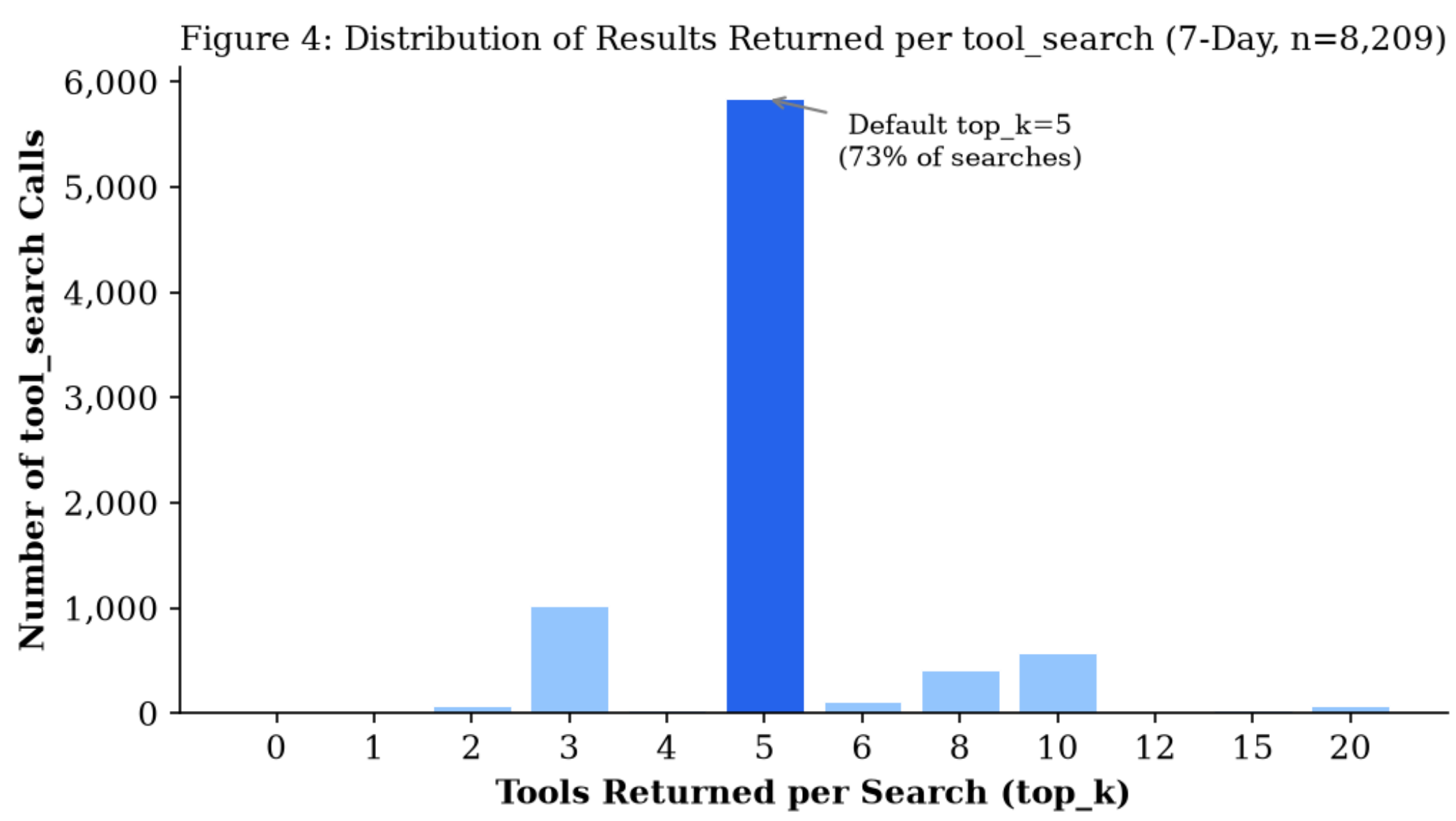

*Figure 4: Distribution of tools returned per tool_search call (7-day, n = 8,209). The default top_k=5 accounts for 73% of searches; clients requesting top_k=3, 8, or 10 represent intentional lower-precision or higher-recall configurations.*

## 8 Discussion

### 8.1 The Meta-Tool Protocol Pattern

Exposing tool_search itself as an MCP tool is a key design decision with broad implications. It means that any MCP-compliant client immediately gains semantic tool discovery without client-side modifications. In production, SCOUT has been validated across six MCP clients: Claude Code, Claude Desktop, ChatGPT (Web and Desktop), GitHub Copilot (VS Code), Cursor, and OpenAI Codex CLI spanning multiple LLM providers and interface types. The two-tool interface (tool_search → execute_tool) creates a natural workflow that all validated clients follow reliably, guided entirely by the tool description text. The tool descriptions are loaded from template files at server startup, allowing discovery guidance to be updated without code changes.

### 8.2 Limitations

Several limitations merit acknowledgment. First, tool descriptions authored by MCP server developers vary significantly in quality and verbosity; poorly described tools may be systematically under-retrieved regardless of retrieval modality, and the system has no mechanism to detect or compensate for this. Second, the system does not currently support feedback-loop learning. Successful tool invocations are not used to refine embeddings or boost retrieval scores for future queries. Third, the sync cron runs every 15 minutes, meaning tool schema changes can take up to 15 minutes to propagate into the search index; time-sensitive schema updates require a manual force-refresh. Fourth, the top-k parameter must be tuned manually by the caller. The system provides no automatic calibration based on query complexity, tool overlap, or result confidence, which may lead to under-retrieval for broad queries or wasted context for narrow ones.

### 8.3 Future Directions

Several directions are being actively explored. Feedback-weighted re-ranking boosting tools that are frequently invoked successfully for similar queries could improve retrieval precision over time without retraining the embedding model. Multi-vector representations (one dense vector per semantic aspect of a tool) may improve disambiguation for tools with broad or overlapping parameter spaces. Automatic top-k calibration dynamically adjusting the number of returned tools based on query complexity and result confidence would reduce the burden of manual parameter tuning.

## 9 Conclusion

We have presented the design and implementation of a production semantic tool discovery system integrated into PayPal's AI Proxy platform. The system addresses the scalability limitations of static tool provisioning in enterprise MCP deployments through five complementary mechanisms: hybrid BM25 + dense retrieval with RRF fusion, 3,072-dimensional embeddings, scope-based authorization filtering, zero-downtime index management, and graceful degradation to full tool lists under infrastructure failure.

The meta-tool protocol exposing discovery and execution as standard MCP tools enables seamless adoption by any MCP-compatible client without client-side modifications. This design embeds semantic tool discovery as a first-class capability within the MCP ecosystem rather than an external pre-processing layer, making the approach composable with future MCP protocol extensions.

## References

[1] Anthropic. Model Context Protocol Specification, 2024. https://spec.modelcontextprotocol.io

[2] T. Schick, J. Dwivedi-Yu, R. Dessì, R. Raileanu, M. Lomeli, L. Zettlemoyer, N. Cancedda, T. Scialom. Toolformer: Language Models Can Teach Themselves to Use Tools. NeurIPS 2023.

[3] S. Yao, J. Zhao, D. Yu, N. Du, I. Shafran, K. Narasimhan, Y. Cao. ReAct: Synergizing Reasoning and Acting in Language Models. ICLR 2023.

[4] T. Significant-Gravitas. AutoGPT: An Autonomous GPT-4 Experiment. GitHub Repository, 2023. https://github.com/Significant-Gravitas/AutoGPT

[5] B. Xu, Z. Peng, B. Lei, S. Mukherjee, Y. Liu, D. Xu. ReWOO: Decoupling Reasoning from Observations for Efficient Augmented Language Models. arXiv:2305.18323, 2023.

[6] P. Lewis, E. Perez, A. Piktus, F. Petroni, V. Karpukhin, N. Goyal, H. Küttler, M. Lewis, W.-T. Yih, T. Rocktäschel, S. Riedel, D. Kiela. Retrieval-Augmented Generation for Knowledge-Intensive NLP Tasks. NeurIPS 2020.

[7] G. V. Cormack, C. L. Clarke, S. Buettcher. Reciprocal Rank Fusion Outperforms Condorcet and Individual Rank Learning Methods. SIGIR 2009.

[8] J. Wang, X. Yi, R. Guo, H. Jin, P. Xu, S. Li, X. Wang, X. Guo, C. Li, X. Xu, K. Yu, Y. Yang, J. Zhou, J. Tang. Milvus: A Purpose-Built Vector Data Management System. SIGMOD 2021.

[9] N. F. Liu, K. Lin, J. Hewitt, A. Paranjape, M. Bevilacqua, F. Petroni, P. Liang. Lost in the Middle: How Language Models Use Long Contexts. Transactions of the Association for Computational Linguistics, 2024.

[10] Yujia Qin, Shihao Liang, et al. ToolLLM: Facilitating Large Language Models to Master 16000+ Real-world APIs. arXiv:2307.16789, 2023.

[11] S. G. Patil, T. Zhang, X. Wang, J. E. Gonzalez. Gorilla: Large Language Model Connected with Massive APIs. arXiv:2305.15334, 2023.

[12] Y. Luan, J. Eisenstein, K. Toutanova, M. Collins. Sparse, Dense, and Attentional Representations for Text Retrieval. Transactions of the Association for Computational Linguistics, 2021.

[13] S. Sen, A. Kasturi, E. Lumer, A. Gulati, V. K. Subbiah. Is Grep All You Need? How Agent Harnesses Reshape Agentic Search. arXiv:2605.15184, May 2026.

[14] V. Karpukhin, B. Oguz, S. Min, P. Lewis, L. Wu, S. Edunov, D. Chen, W. Yih. Dense Passage Retrieval for Open-Domain Question Answering. EMNLP 2020.

[15] N. Thakur, N. Reimers, A. Rücklé, A. Srivastava, I. Gurevych. BEIR: A Heterogenous Benchmark for Zero-shot Evaluation of Information Retrieval Models. NeurIPS 2021.